\documentclass[a4paper,11pt]{article}
\pdfoutput=1

\usepackage{jheppub}
\usepackage[T1]{fontenc}
\usepackage{amssymb}
\usepackage{graphicx}
\usepackage{amsmath}
\usepackage{tensor}
\usepackage{hyperref}
\usepackage{epstopdf}
\usepackage{extarrows}
\usepackage{xcolor}
\usepackage{multirow}

\newcommand{\td}{\text{d}}
\def\I {\hat{\mathbb{I}}}
\def\x {\hat{x}}
\def\p {\hat{p}}

\def\O {\hat{O}}
\def\U {{\hat{U}}}
\def\W {{\hat{W}}}

\def\Tr {{\text{Tr}}}

\def\C {\mathcal{C}}
\def\F {\tilde{F}}

\def\H {{\cal H}}
\def\bH{\bar{\H}}

\begin{document}
\title{Complexity of operators generated by quantum mechanical Hamiltonians}

\author[a]{Run-Qiu Yang}
\author[b]{and Keun-Young Kim}

\emailAdd{aqiu@kias.re.kr}
\emailAdd{fortoe@gist.ac.kr}

\affiliation[a]{Quantum Universe Center, Korea Institute for Advanced Study, Seoul 130-722, Korea}
\affiliation[b]{ School of Physics and Chemistry, Gwangju Institute of Science and Technology,
Gwangju 61005, Korea
}

\abstract{
We propose how to compute the complexity of operators generated by Hamiltonians in quantum field theory (QFT) and quantum mechanics (QM).
The Hamiltonians in QFT/QM and quantum circuit have a few essential differences, for which we introduce new principles and methods for complexity. We show that the complexity geometry corresponding to one-dimensional quadratic Hamiltonians is equivalent to AdS$_3$ spacetime.
Here, the requirement that the complexity is nonnegative corresponds to the fact that the Hamiltonian is lower bounded and the speed of a particle is not superluminal.
 Our proposal proves the complexity of the operator generated by a free Hamiltonian is zero, as expected. By studying a non-relativistic particle in compact Riemannian manifolds we find the complexity is given by the global geometric property of the space. In particular, we show that in low energy limit the critical spacetime dimension to ensure the `nonnegative' complexity is the 3+1 dimension.
}

\maketitle

\section{Introduction}

The recent developments of the quantum information theory and holographic duality show that some concepts from quantum information are useful in understanding foundations of gravity and quantum field theory (QFT). One of such concepts is the entanglement entropy, from which the spacetime geometry may emerge (e.g., see Refs.~\cite{PhysRevLett.96.181602,Nishioka:2009un,VanRaamsdonk:2010pw,Nozaki:2012zj,Lin:2014hva,Hayden:2016cfa}).  The ``complexity''  is another important concept since it may have a relation with physics {\it inside} black hole horizon~\cite{Harlow:2013tf,Stanford:2014jda,Susskind:2014rva,Stanford:2014jda,Brown:2015bva,Brown:2015lvg,Czech:2017ryf}.

  Note that the deep understandings of the entanglement entropy in the context of holography was possible because it is well-defined in both gravity and QFT. While there has been much progress for the complexity in gravity side \footnote{For example, see~\cite{Susskind:2014rva, Stanford:2014jda, Roberts:2014isa,Brown:2015bva,Brown:2015lvg,Cai:2016xho,Lehner:2016vdi,Chapman:2016hwi,Carmi:2016wjl,Reynolds:2016rvl,Kim:2017lrw,Carmi:2017jqz,Kim:2017qrq,Swingle:2017zcd}.}, even the precise definition of the complexity in QFT is still not complete, which is an essential question.

 The complexity is originally defined in quantum computation theory. Suppose a unitary operator $\U$ is simulated by a quantum circuit, which can be constructed by combining universal elementary gates by many ways. The complexity of $\U$ is defined by the minimal number of required gates to build $\U$. This definition by ``counting'' is based on the finite and discrete systems. However, to define the complexity in QFT or quantum mechanics (QM) we have to deal with infinite and continuous systems: we cannot ``count''.

 Recently, there have been a few attempts to generalize the concept of complexity of discrete quantum circuit to continuous systems, e.g., ``complexity geometry''~\cite{Susskind:2014jwa,Brown:2016wib,Brown:2017jil} based on ~\cite{Nielsen1133,Nielsen:2006:GAQ:2011686.2011688,Dowling:2008:GQC:2016985.2016986}, Fubini-study metric~\cite{Chapman:2017rqy}, and path-integral optimization~\cite{Caputa:2017urj,Caputa:2017yrh,Bhattacharyya:2018wym,Takayanagi:2018pml}. See also \cite{Hashimoto:2017fga,Hashimoto:2018bmb,Flory:2018akz}. In particular, the complexity geometry is
 the most studied idea (for example see~Refs.~\cite{Jefferson:2017sdb,Yang:2017nfn,Reynolds:2017jfs,Kim:2017qrq,Khan:2018rzm,Hackl:2018ptj,Yang:2018nda,Yang:20180919,Alves:2018qfv,Magan:2018nmu,Caputa:2018kdj,Camargo:2018eof,Guo:2018kzl,Bhattacharyya:2018bbv, Jiang:2018gft,Camargo:2018eof,Chapman:2018hou,Ali:2018fcz}).
      In essence, one introduces a certain right-invariant Riemannian 
      geometry where every points corresponds to a unitary operator. The continuous versions of quantum circuits of a given $\U$ are identified with various curves connecting the identity and $\U$ in the geometry. The complexity is defined by the minimal length of those curves.

 This idea of the complexity geometry of unitary operators is very attractive but current studies have two important shortcomings yet.  First,
 the geometry is not determined by some physical principles, but given by hand. It is acceptable for quantum circuit problems since we can design the circuit as we want. However, in general QFT/QM, there must be constraints given by nature not by our hands.
 Second, it is only valid for SU($n$) operators with finite $n$. It is enough for finite qubit systems, but to deal with the operators generated by  general QFT/QM systems such as $H=p^2/2m + V(x,p)$ we need to develop the formalism for the Hamiltonian with an infinite dimensional Hilbert space.

In this paper, we propose how to remedy these shortcomings by generalizing the complexity geometry of finite qubit systems to general QFT/QM.  In particular, we take into account the fact that the generating functional in QFT/QM plays a crucial role contrary to quantum circuits. We also, for the first time, note the importance of the lower boundedness of the Hamiltonian in QFT/QM, in the context of complexity. As a result, we will uncover novel interesting results which cannot be simply inferred from finite qubit systems.

\section{Complexity of unitary operators generated by Hamiltonians}    \label{regulC1}

\subsection{Overview on Nielsen's right-invariant complexity geometry}\label{overviweC}
Let us first review basic ideas of right-invariant complexity geometry for SU(n) groups based on~\cite{Nielsen1133,Nielsen:2006:GAQ:2011686.2011688,Dowling:2008:GQC:2016985.2016986}.
We consider the space of operators in SU($n$) group with finite $n$. Suppose that a curve ($c(s)\in$SU($n$)) is generated by a generator $H(s)$ as follows.
\begin{equation}\label{defRH1}
c(s) = \overleftarrow{\cal{P}}e^{\int_0^s i H(\tilde{s}) \td \tilde{s}} \quad  \mathrm{or} \quad  \dot{c}(s)=iH(s)c(s)\,.
\end{equation}
%
%
We assume that the line element of this curve is given by a certain function of a generator only:
\begin{equation}\label{rmetric1}
  \td l= \F(H(s)) \td s :=\sqrt{\tilde{g}(H(s),H(s))} \td s\,,
\end{equation}
{where $\tilde{g}(\cdot,\cdot)$ is required to be an inner product for the Lie algebra  $\mathfrak{su}(n)$ and a quadratic in $H$.} {In fact, it is possible to consider a more general inner product than a quadratic form. This will yield a general Finsler geometry rather than a Riemannian one~\cite{Yang:2018nda}. In this paper we assume that the complexity geometry is a Riemannian geometry.} We call $\F(H)$ the {\it norm} of $H$.
%
%
%
%
%
In bases $\{e_I\}$, $H=e_IY^I$ and the metric components can be expressed as
\begin{equation}\label{defmetricg}
  \tilde{g}_{IJ}=\frac12\partial^2\F(H)^2/(\partial Y^I\partial Y^J)\Leftrightarrow\F(H)^2=\tilde{g}_{IJ}Y^IY^J\,.
\end{equation}
Note that giving the norm $\F$ is equivalent to giving a metric $\tilde{g}_{IJ}$ under a bases.

To obtain the metric in the group manifold with the coordinate $X^I$, the metric needs to be transformed by a coordinate transformation
\begin{equation} \label{groupmetric}
g_{IJ}(X) = \tilde{g}_{KL} M^K_I(X) M^L_J(X) \,,
\end{equation}
where the transformation matrix is defined as $Y^I(s) \td s = M^I_K (X) \td X^K$. (See appendix \ref{app2} for a concrete example.)
%
%
The complexity of an operator $\hat{W}(s) :=\overleftarrow{\cal{P}}e^{\int_0^s i H(\tilde{s}) \td \tilde{s}}$
, denoted by $\C(\hat{W}(s))$, is defined by the minimal length of all curves which connect $\hat{W}(s)$ to identity:
\begin{equation}\label{defmin}
\C(\hat{W}(s))) = \min \int_0^s \F(H(\tilde{s})) \td \tilde{s} \,,
\end{equation}
where $H(\tilde{s})$ satisfy $\hat{W}(s)=\overleftarrow{\cal{P}}e^{\int_0^s i H(\tilde{s}) \td \tilde{s}}$.

After we obtain the complexity for all operators in the SU(n) group based on Eq.~\eqref{defmin}, the complexity between two pure quantum states in an $n$-dimensional Hilbert space can be expressed as the following optimal problem
\begin{equation}\label{Cforsates}
  \C(|\psi_1\rangle,|\psi_2\rangle)=\min\left\{\C(U)~|~\forall~\U\in\mathcal{O},~~|\psi_2\rangle=\U|\psi_1\rangle\right\} \,,
\end{equation}
{where the unitary operator may belong to some restricted set $\mathcal{O}$, which is a subgroup of SU($n$) group and depends on detailed physical problems.}
Thus, the norm $\F$ plays a central role when we analyse the complexity in quantum systems. Once we obtain the norm $\F$, the metric in the SU($n$) group (and its any subgroup) is computed.  By this metric, the minimal geodesic length connecting the identity and the target operator, which is nothing but the complexity of the operator, is computed. The complexity between two states is the minimal complexity of the operators shown in Eq.~\eqref{Cforsates}. In this paper, we will only focus on the complexity of unitary operators.

Note that the complexity is right-invariant, because $H$ itself is invariant under the right-translation $c \rightarrow c \hat{x}$ for $\forall \hat{x} \in $ SU($n$). However, for a left translation $c(s)\mapsto\hat{x} c(s)$,  the generator will be transformed as
$$H(s)\mapsto\x H(s)\x^{\dagger}\,,$$
which is different from $H(s)$ in general. If there is no additional symmetry, $\F(H)\neq \F(\x H\x^{\dagger})$, the complexity is not left-invariant but only right-invariant.




\subsection{Bi-invariant complexity geometry}
Nielsen's (only) right-invariant complexity is a good tool for the studies on   quantum computation and quantum circuit systems. Many recent works such as~\cite{Nielsen1133,Nielsen:2006:GAQ:2011686.2011688,Dowling:2008:GQC:2016985.2016986} and \cite{Brown:2017jil,Jefferson:2017sdb,Chapman:2017rqy,Kim:2017qrq,Camargo:2018eof} try to generalize this idea to the studies on QFT/QM. These works assume that the complexity is only right-invariant. 
However,  if the complexity in QFT/QM is only right-invariant, we find that there are some issues which cannot be reconciled with basic principles or symmetries of QFT/QM.
These subtleties have been discussed in detail in Refs.~\cite{Yang:2018nda,Yang:20180919}. For readers' convenience, in appendix \ref{unitrary1}, we make a brief review only for two arguments among various arguments in Refs.~\cite{Yang:2018nda,Yang:20180919}. See also section \ref{uni123}.

A punchline of appendix \ref{unitrary1} and Refs.~\cite{Yang:2018nda,Yang:20180919} is that the complexity in QFT/QM indeed needs to have a symmetry
\begin{equation}\label{untrary-inv}
  \F(H)=\F(\U H\U^\dagger)\,,
\end{equation}
which we will call a unitary invariance.
The right-invariance and unitary invariance~\eqref{untrary-inv} together imply the complexity in QFT/QM should be also left-invariant, so {\it bi-invariant}. One very quick argument to support this unitary invariance is as follows.

As a generator $H$ in \eqref{defRH1} let us consider a Hamiltonian of QFT/QM.  Recall that $H$ and $\U H \U^{\dagger}$ give the same generating functional so they are physically equivalent. Therefore, if the complexity in QFT/QM is an observable or a physical quantity yielding observables
it is natural that the complexity of the unitary operators generated by $H$ and $\U H \U^{\dagger}$ are also same, which amounts to Eq~\eqref{untrary-inv}. For clarify our point on Eq.~\eqref{untrary-inv}, we would like to emphasize again that the only right-invariance is perfectly fine for quantum circuits. What we want to highlight is that there is something more to consider for general QFT/QM systems which are different from quantum circuits.

%
%
%

The bi-invariance plays a very important role in determining a complexity geometry and a geodesic in there, based on two important mathematical results. First, the theory of Lie algebra proves that the Riemannian metric $\tilde{g}(\cdot,\cdot)$ is uniquely determined by the Killing form up to a constant factor $\lambda > 0$~\cite{Alexandrino2015}
\begin{equation}\label{GbiFs1}
  \tilde{g}(H,H)=\F(H)^2 = \lambda^2  \Tr(HH^\dagger)= \lambda^2  \Tr(H^2) \,.
\end{equation}
%
Second, the geodesic in a bi-invariant metric is given by a {\it constant}
 generator~\cite{Alexandrino2015}, say $\bar{H}$,  so Eq. \eqref{defmin} yields\footnote{For an operator $\exp(i\bar{H})$ with a given $\bar{H}$, there may be many $H_k$ satisfying $\exp(iH_k)=\exp(i\bar{H})$.  The complexity is given by the minimal $\F(H_k)$ among all $H_k$. Thus, the complexity of the operator $\exp(iHt)$ may stop growing after large time $t$~\cite{Yang:20180919}. }
\begin{equation}\label{CforHs1}
\C(\O)=\F(\bar{H})~~\text{with}~\exp(i\bar{H})=\O\,.
\end{equation}
Note that the work of finding geodesic is greatly simplified due to bi-invariance.

In this paper, we will discuss the complexity in the current frameworks of QM and QFTs and so the complexity is bi-invariant. Let us now consider an operator generated by a physical Hamiltonian, denoted by $\H$.  In general  $\H$ has a {\it nonzero trace} and its Hilbert space may be {\it infinite} dimensional so Eq.~\eqref{GbiFs1} needs to be generalized accordingly.

First, to deal with the Hamiltonian of nonzero trace, we define a ``mean value'' $\bH$ by
\begin{equation}\label{meanHs1}
  \Tr(\H- \bH \I) = 0\,, \quad  \I ={\mathrm{identity}} \,,
\end{equation}
such that $(\H-\bH \I)\in\mathfrak{su}(n)$.
Because $U(1)$ is just a phase transformation, $(\H-\bH\I)$ and $\H$ generate equivalent transformations and they should give the same complexity.  Thus, we may use Eq.~\eqref{GbiFs1} for the norm of $\H$
\begin{equation}\label{formforFs1b}
  \F(\H) = \lambda \sqrt{\Tr[(\H-\bH\I)^2]}  \,. 
\end{equation}
%
%

However, if the Hilbert space of $\H$ is infinite dimensional the trace in Eq.\eqref{meanHs1} and \eqref{formforFs1b} are divergent \footnote{In a mathematical jargon, the Hamiltonian may not be in the {\it trace class}.} so we need to renormalize it. We will show how to do it for two important cases: a quadratic Hamiltonian in one dimensional space and a non-relativistic particle in compact Riemannian manifolds.

\section{AdS$_3$ spacetime as a complexity geometry}\label{quadH}
Let us consider a general quadratic Hamiltonian in one-dimensional space,
\begin{equation}\label{quadraticH1}
  \H=\frac{Y^+}2\x^2+\frac{Y^-}2\p^2+\frac{Y^0}{4}(\x\p+\p\x) \,.
\end{equation}
Here, we want to emphasize that the $\{Y^I\}$ should satisfy the constraints
\begin{equation}\label{phyquadcond0}
  4Y^+Y^--(Y^0)^2 \ge 0~~~\text{and}~~Y^{\pm}\geq0\,,
\end{equation}
so that the Hamiltonian is bounded below. It can be seen from Eq. \eqref{relc1c2} and {will play an important role in the comparison of our complexity geometry with AdS spacetime in Eq.~\eqref{metricSU112spacetime}.  This boundedness condition} is another difference between QFT/QM and quantum circuits. The Hermit operators in quantum circuits have finite number of eigenvalues so they are always bounded below. However, the Hermit operators in QFT/QM may not be bounded below in general, so we should be careful in defining parameter ranges.

To compute the norm of $\H$ we make a canonical transformation $(x,p)\rightarrow(x',p')$ as follows,
\begin{equation}\label{quadraticH3}
  \H=\frac{c_1}2\hat{x}'^2+\frac{c_2}2\hat{p}'^2\,,
\end{equation}
where
\begin{equation}\label{relc1c2}
  c_1+c_2=Y^++Y^-\,, \quad c_1c_2=Y^+Y^--(Y^0)^2/4\,.
\end{equation}
%
The eigenvalues of $\H-\bH\I$ read
\begin{equation}\label{eigenEn}
  E_n=\omega(n+1/2)-\bH,~~~n=0,1,2,\cdots\,
\end{equation}
with $\omega=\sqrt{Y^+Y^--(Y^0)^2/4}$. {The norm of $\H$, Eq. \eqref{formforFs1b}, will be divergent and we need to consider its proper renormalization. }

In general, we can define the $\zeta$-function for a positive definite Hermit operator $\cal{O}$ with eigenvalues $\{ E_n\}$ as analytic continuation of the following sum
\begin{equation}
  \zeta_{\cal{O}}(s):=\sum_{n=0}^\infty\frac1{E_n^s}\,.
\end{equation}
In particular, for the operator ${\cal{O}} = \H-\bH\I$ with $\H$ in~\eqref{quadraticH3}, we have
\begin{equation}\label{defzeta4}
  \zeta_{\H-\bH\I}(s)=\sum_{n=0}^\infty\frac{\omega^{-s}}{(n+1/2-\bH/\omega)^s}=\omega^{-s}\zeta(s,1/2-\bH/\omega)\,,
\end{equation}
where $\zeta(s,q)$ is the Hurwitz-$\zeta$ function, which is defined as the analytic continuation of the following sum~\cite{elizalde2012ten}
\begin{equation}\label{huriwitz}
  \zeta(s,q)=\sum_{n=0}^\infty\frac1{(n+q)^s}\,.
\end{equation}
We may define the renormalized trace as
\begin{equation}\label{defzetaTr1}
  \Tr_{\text{re}}(\H-\bH\I)=\zeta_{\H-\bH\I}(-1)\,,
\end{equation}
and the renormalized norm squared as
\begin{equation}\label{defzetaTr3}
  \F^2_{\text{re}}(\H)=\lambda^2 \Tr_{\text{re}}[(\H-\bH\I)^2]= \lambda^2 \zeta_{\H-\bH\I}(-2)\,.
\end{equation}

First, by Eq.~\eqref{meanHs1}, Eq.~\eqref{defzetaTr1} determines $\bH$:
%
%
%
\begin{equation}\label{valueunderlineHzeta}
  \zeta(-1,-\bH/\omega+1/2)=0~~\Rightarrow~~\frac{\bH^2}{\omega^2}-\frac1{12}=0\,,
\end{equation}
which gives two solutions $\bH=\pm\omega/(2\sqrt{3})$.
%
%
Thus, the Eq.~\eqref{defzetaTr3} becomes
\begin{equation}\label{trH2quadratic}
  \F^2_{\text{re}}(\H) =
  \lambda^2\omega^2\zeta\left(-2,\pm\frac1{2\sqrt{3}}+ \frac12\right)=\pm\frac{\sqrt{3}\lambda^2}{108}\omega^2\,.
\end{equation}
Because $\F^2_{\text{re}}(H)$ should be nonnegative when coefficients $Y^I$ satisfy Eq.~\eqref{phyquadcond0}, we have to choose $\bH=-\omega/(2\sqrt{3})$.
%
Defining the parameter $\lambda^2=108\lambda^2_0/\sqrt{3}$ we obtain
\begin{equation} \label{F211}
F^2_{\text{re}}(\H)=\lambda_0^2\omega^2 = \lambda_0^2 \left(Y^+Y^--(Y^0)^2/4 \right) \,.
\end{equation}
%

Indeed, our result \eqref{F211} is consistent with the group theoretic consideration. The Hamiltonian \eqref{quadraticH1} is written as $\H=Y^Ie_I$ with
\begin{equation}\label{threeeforSU11}
  e_+=\frac{1}{2}\x^2\,, \quad e_-=\frac{1}{2}\p^2\,,\quad e_0=\frac{1}4(\x\p+\p\x)\,,
\end{equation}
where $\{ie_I\}$ consists of the $\mathfrak{su}(1,1)$ Lie algebra \footnote{This is also the symplectic Lie algebra $\mathfrak{sp}(2,\mathbb{R})$, the special linear  Lie algebra $\mathfrak{sl}(2,\mathbb{R})$ and Lorentz Lie algebra $\mathfrak{so}(2,1)$.}
\footnote{The complexity related to  $\mathfrak{su}(1,1)$ algebra has been studied in Refs.~\cite{Jefferson:2017sdb,Chapman:2017rqy,Kim:2017qrq,Camargo:2018eof}, in only right-invariant geometry and without the restriction~\eqref{phyquadcond0}.}
\begin{equation}\label{su11algebra}
  [ie_0,ie_\pm]=\pm ie_\pm\,,  \quad [ie_-,ie_+]=2ie_0 \,.
\end{equation}
%
As the $\mathfrak{su}(1,1)$ is simple, its bi-invariant metric $\tilde{g}(\cdot,\cdot)$ is proportional to the Killing form~\cite{Alexandrino2015}. In other words, $\tilde{g}$ can be computed simply by the structure constants of the Lie algebra. As a result
\begin{equation}\label{metricfromB1}
  \tilde{g}_{IJ}Y^IY^J \sim 4Y^+Y^--(Y^0)^2\,,
\end{equation}
which serves as a nice consistent check for our method. Even more interestingly, in section \ref{3rdway}, we provide an alternative independent method to obtain metric~\eqref{metricfromB1} {\it without assuming the symmetry~\eqref{untrary-inv}}. This is important because it also justifies the unitary symmetry Eq. \eqref{untrary-inv} or the bi-invariance in a nontrivial way. 


Let us consider the case, $Y^+ = Y^0 = 0$ and $Y^-> 0$, which corresponds to the free particle case.
From \eqref{F211} we find that
\begin{equation}\label{normfreeH21}
\F_{\text{re}}(\H)=0\,,
\end{equation}
which is the first explicit realization of the fact~\cite{Brown:2015lvg} (the Sec.~VIII B 4) that the complexity of the operator generated by a free Hamiltonian is zero because a free Hamiltonian cannot mix information across its degrees of freedom.

\begin{figure}
  \centering
  \includegraphics[width=.45\textwidth]{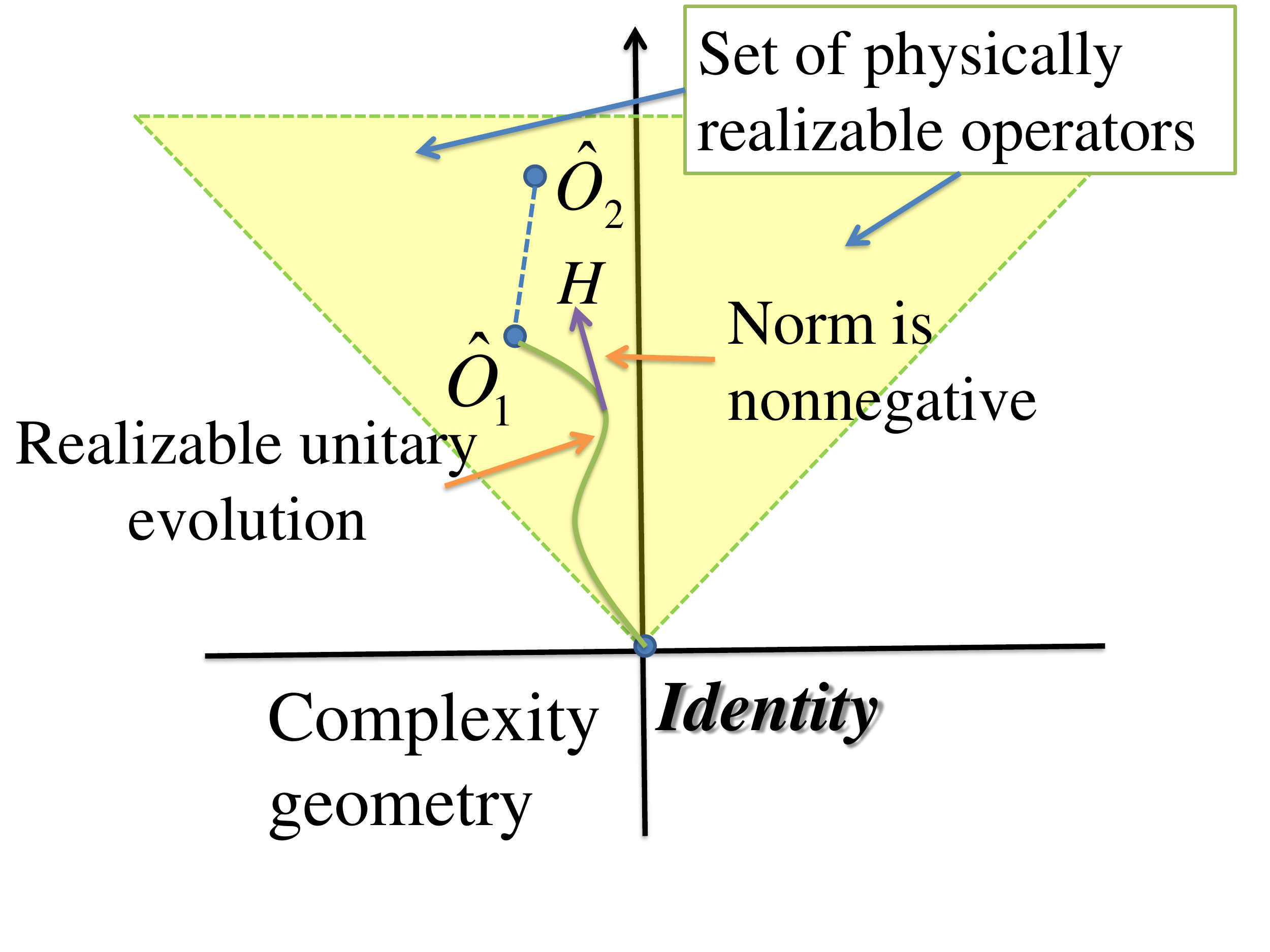}
  \includegraphics[width=.45\textwidth]{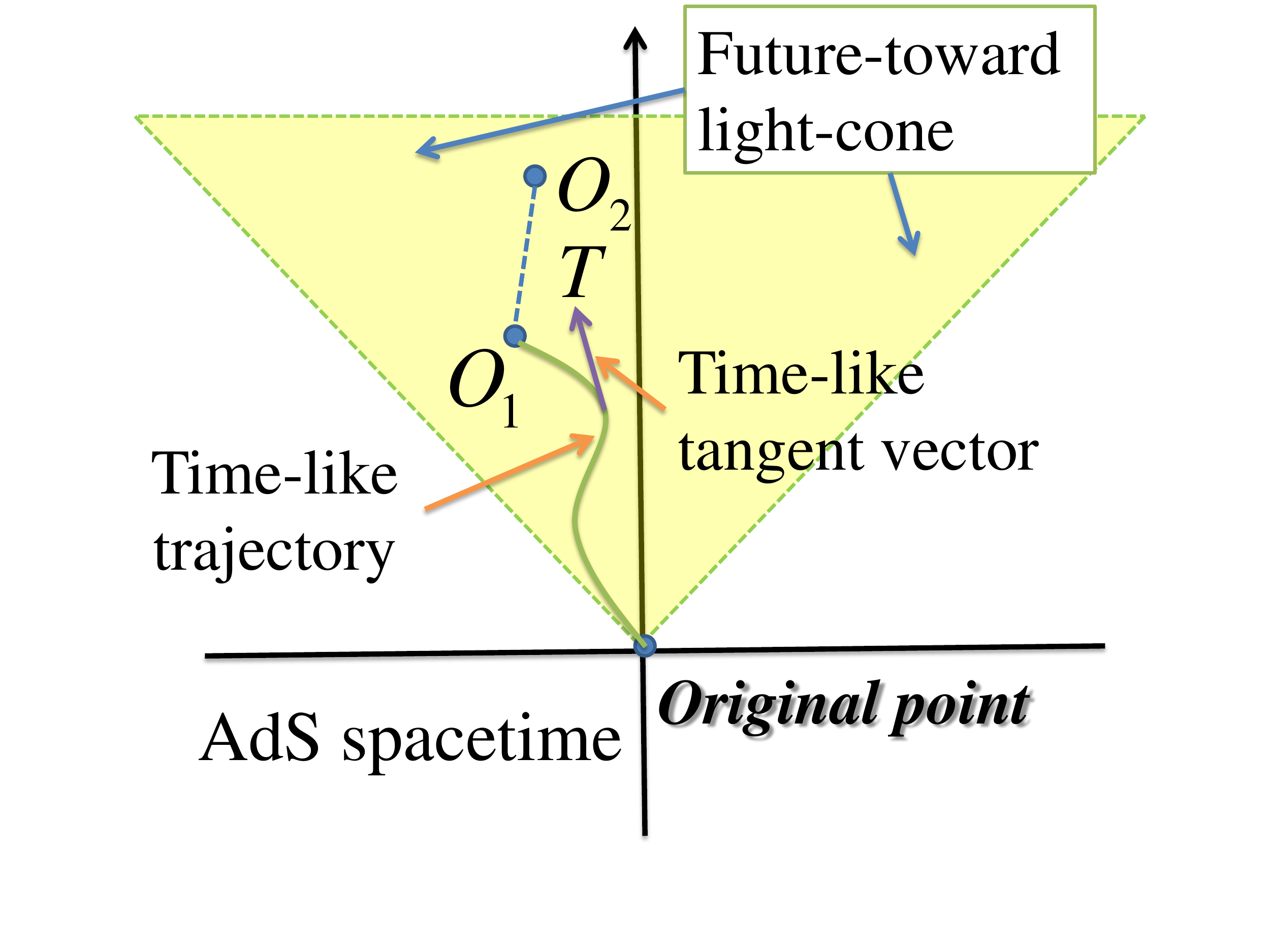}
  \caption{Schematic diagram on the similarities between the complexity geometry of the quadratic Hamiltonian (left panel) and the spacetime geometry of AdS$_3$ (right panel).}\label{RelCAdS3}
\end{figure}
\begin{table*}
\centering
\begin{tabular}{c|c}
  \hline
  Complexity geometry & Spacetime geometry \\
   \hline
   Unitary operators generated &\multirow{2}{*}{Points in AdS$_{3}$ spacetime}\\
   by quadratic Hamiltonian& \\
   \hline
   Norm is nonnegative & Speed of particle cannot be superluminal\\
   \hline
   \multirow{2}{*}{Curves with positive complexity}& Trajectories with future-toward\\
   &time-like tangent vectors\\
   \hline
   Set of physically realizable unitary operators& Future-toward light-cone\\
  \hline
\end{tabular}
\caption{Comparison between the complexity geometry of SU(1,1) group and spacetime geometry of AdS$_3$. }\label{tab1}
\end{table*}

By parameterizing the SU(1,1) group as
\begin{equation}\label{parameterU112}
  \U(y, z,u)=\exp(iy e_0)\exp(iz e_+)\exp(iue_-)\,,
\end{equation}
we obtain the metric in the group manifold (see Eq.~\eqref{groupmetric} and appendix~\ref{app2} for details)
\begin{equation}\label{zetametricSU11}
  \td l^2=\lambda_0^2(-\td y^2/4+z\td y\td u+\td z\td u)\,.
\end{equation}
%
As the signature of this metric is $(-,-,+)$, it is interesting to define a ``spacetime interval'' $\td s^2=-\td l^2<0$
\begin{equation}\label{metricSU112spacetime}
  \td s^2=g_{IJ}^{(\text{st})}\td x^I\td x^J :=\lambda_0^2(\td y^2/4-z\td y\td u-\td z\td u)\,,
\end{equation}
which yields the Riemann tensor $R_{IJKL}^{(\text{st})}$
\begin{equation}\label{RicciCootonSU11}
 R_{IJKL}^{(\text{st})}=-\frac1{\lambda_0^2}(g_{IK}^{(\text{st})}g_{JL}^{(\text{st})}-g_{IL}^{(\text{st})}g_{JK}^{(\text{st})})\,.
\end{equation}
%
This means that the AdS$_3$ spacetime with the AdS radius $\ell_{\text{AdS}}=\lambda_0$  emerges as our complexity geometry!
We find some interesting correspondences between the complexity geometry of SU(1,1) group and AdS$_3$ spacetime. For examples, the {\it physical} Hamiltonians with \eqref{phyquadcond0} correspond to the {\it time-like} or {\it null} tangent vectors in AdS$_3$ spacetime. In Fig.~\ref{RelCAdS3} and Tab.~\ref{tab1}, we make more comparisons between the  complexity geometry of SU(1,1) generated by quadratic Hamiltonian and spacetime geometry of AdS$_3$. It is interesting that three basic requirements in physics, i) the complexity should be nonnegative, ii) the Hamiltonian should be bounded below and iii) the speed of particles cannot be superluminal, are three different facets of the same in this model.

\subsection{Alternative method to obtain the complexity geometry for quadratic Hamiltonians}\label{3rdway}
In this subsection, we provide alternative method to obtain the complexity geometry~\eqref{metricfromB1} for quadratic Hamiltonians.  Here, we do not assume  the complexity geometry is bi-invariant but only assume right-invariant. Just by using  $U(1)$ gauge symmetry and canonical transformation, we can determine the unique complexity geometry. Furthermore, the result shows that the complexity geometry of general quadratic Hamiltonians must be bi-invariant and satisfy a symmetry $\F(H)=\F(\U H\U^\dagger)$. We also provide another example to show why symmetry $\F(H)=\F(\U H\U^\dagger)$ 
is necessary.

Let us consider a Hamiltonian of a non-relativistic particle with potential $V(\vec{\x})=k\vec{\x}^2/2$ and a magnetic vector potential $\vec{A}(\vec{\x})$,
\begin{equation}\label{HamilA1}
  \H=\frac1{2m}[\vec{\p}-q\vec{A}(\vec{\x})]^2+\frac{k}2\vec{\x}^2\,,
\end{equation}
where $q$ is the charge carried by the particle.
If we take a pure gauge, $\vec{A}(\vec{\x})=\nabla\phi(\vec{\x})$ for an arbitrary scalar function $\phi(\vec{\x})$, the Hamiltonian
\begin{equation}\label{HamilA2}
  \H_\phi=\frac1{2m}[\vec{\p}-q\nabla\phi(\vec{\x})]^2+\frac{k}2\vec{\x}^2\,,
\end{equation}
and the harmonic oscillator Hamiltonian
\begin{equation}\label{HamilA3}
  \H_0=\frac1{2m}\vec{\p}^2+\frac{k}2\vec{\x}^2\,,
\end{equation}
describe the same physical system. Thus, they should give the same complexity and
\begin{equation}\label{gaugesy1}
  \F(\H_0)=\F(\H_\phi) \,,
\end{equation}
for all $\phi(\vec{\x})$. For the Hamiltonian~\eqref{HamilA3}, we can take a canonical transformation $(\vec{\x},\vec{\p})\rightarrow(\gamma\vec{\x},\gamma^{-1}\vec{\p})$, which leads to
\begin{equation}\label{HamilA3b}
  \H_0(\gamma)=\frac1{2m\gamma^2}\vec{\p}^2+\frac{k\gamma^2}2\vec{\x}^2\,.
\end{equation}
Because this transformation is canonical, $\H_0(\gamma)$ and $\H_0$ describe the same physical system so
\begin{equation}\label{gaugesy2}
  \F(\H_0)=\F(\H_0(\gamma))\,,
\end{equation}
for all $\gamma\neq0$.

Let us first consider what we can obtain from the symmetry~\eqref{gaugesy1}. For $\phi(\vec{\x})=-aq^{-1}m\vec{\x}^2$ with an  arbitrary real number $a$ we obtain
\begin{equation}\label{HamilA2b}
\begin{split}
  \H_\phi&=\H(a,k,m)\\
  :&=\sum_{i=1}^3\frac1{m}\frac{\p_i^2}2+4a\frac{\x_i\p_i+\p_i\x_i}4+(4a^2m+k)\frac{\x_i^2}2\,.
  \end{split}
\end{equation}
Comparing with the bases defined in Eq.~\eqref{threeeforSU11},
we find  that this Hamiltonian contains the triple copies of $\mathfrak{su}(1,1)$ Lie algebra. The  symmetry~\eqref{gaugesy1} implies
\begin{equation} \label{symm1}
\F(\H(a,k,m))=\F(\H(0,k,m)) \,,
\end{equation}
for arbitrary $a, k$ and $m$. In general, $\F^2(\H) = \tilde{g}_{IJ} Y^I Y^J$, where $\tilde{g}_{IJ}$ is a general metric for $\mathfrak{su}(1,1)$ Lie algebra. In our case,
\begin{equation}\label{normalFA1}
\begin{split}
  &\F^2(\H(a,k,m))=3\left[\tilde{g}_{--}\frac{1}{m^{2}}+\tilde{g}_{00}16a^2+\tilde{g}_{++}(4a^2m+k)^2\right.\\
  &\left.+\tilde{g}_{+-}2\frac{4a^2m+k}m+\tilde{g}_{+0}8a(4a^2m+k)+\tilde{g}_{-0}\frac{8a}m\right]\,.
  \end{split}
\end{equation}
%
The overall coefficient 3 comes from the fact that $\H(a,k,m)$ contains triple copies of $\mathfrak{su}(1,1)$ Lie algebra.
If $k=0$ Eq.~\eqref{symm1} implies
\begin{equation}\label{solveeqgij1}
  \begin{split}
  \tilde{g}_{--}\frac{1}{m^{2}}=&\left[\tilde{g}_{--}\frac{1}{m^{2}}+(2\tilde{g}_{00}+\tilde{g}_{+-})8a^2\right.\\
  &\left.+\tilde{g}_{++}16m^2a^4+\tilde{g}_{+0}32ma^3+\tilde{g}_{-0}\frac{8a}m\right] \,,
  \end{split}
\end{equation}
for all $a$. This means that $\tilde{g}_{++}=\tilde{g}_{+0} = \tilde{g}_{-0}=0$ and
\begin{equation}\label{relggsq1}
  \tilde{g}_{+-}=-2\tilde{g}_{00}\,.
\end{equation}
Thus, Eq.~\eqref{normalFA1} boils down to
\begin{equation}\label{normalFA2b}
  \F^2(\H(a,k,m))=3\left[\tilde{g}_{--}\frac{1}{m^2}+2\tilde{g}_{+-}\frac{k}m\right] \,.
\end{equation}
We see that the gauge symmetry gives us very strong restriction on the metric for quadratic Hamiltonians.

Now let us consider what we can obtain from the symmetry~\eqref{gaugesy2}, which means
\begin{equation}
\F^2(\H(0,k,m) = \F^2(\H(0,k\gamma^2,m\gamma^2) \,.
\end{equation}
By using Eq.~\eqref{normalFA2b}, we have
\begin{equation}\label{normalFA2c}
  \frac{\tilde{g}_{--}}{m^2}+2\tilde{g}_{+-}\frac{k}m=\frac{\tilde{g}_{--}}{(m\gamma^2)^{2}}+2\tilde{g}_{+-}\frac{k}m \,,
\end{equation}
for all $\gamma\neq0$. This implies $\tilde{g}_{--}=0$.

Thus, respecting the symmetries~\eqref{gaugesy1} and~\eqref{gaugesy2} we find that the nonzero components of the metric $\tilde{g}_{IJ}$ are only $\{\tilde{g}_{+-}, \tilde{g}_{00}\}$ and they satisfy Eq.~\eqref{relggsq1}. As a result, for a quadratic Hamiltonian $\H=Y^Ie_I$ with $\{e^I\}$ defined in Eq.~\eqref{threeeforSU11},
%
\begin{equation}\label{F211b}
\F^2(\H)=-g_{00}[4Y^+Y^--(Y^0)^2] \,,
\end{equation}
which is the same as Eq.~\eqref{F211}.

\subsection{Another argument for the unitary invariance}\label{uni123}

Section \ref{3rdway}  also shows that the complexity metric of quadratic Hamiltonians must be bi-invariant! {In this subsection, we will offer more concrete examples to support that  the bi-invariance is natural and necessary  when we consider the ``complexity''  in QFT/QM.}

Let us consider the following Hamiltonian:
\begin{equation}\label{translq1}
  \H(\vec{w})=\frac{\vec{\p}^2}{2m}+\frac{k}{2}\vec{\x}^2+k\vec{w}\cdot\vec{\x}+\frac{k\vec{w}^2}{2}\,.
\end{equation}
The Hamiltonians $ \H(\vec{w})$ and $\H_0$ look different if $\vec{w}\neq0$ and one may think that $ \H(\vec{w})$ and $\H_0$ give different complexities since some additional ``gates'' may be required to realize the additional operator $k\vec{w}\cdot\vec{\x}$ in the Hamiltonian $\H(\vec{w})$. However, Eq. \eqref{translq1} is nothing but
\begin{equation}
\H(\vec{w})=\frac{\vec{\p}^2}{2m}+\frac{k}{2}(\vec{\x}+\vec{w})^2\,,
\end{equation}
which just shifts the coordinate compared to $\H_0$ and does not change physics. Because $\H(\vec{w})$ and $\H_0$ describe the same physical system we should have
\begin{equation}\label{FhFaH}
  \forall\vec{w}\in\mathbb{R}^3,~~~\F(\H_0)=\F(\H(\vec{w}))\,.
\end{equation}
%

In fact, four Hamiltonians $\{\H_0, \H_\phi, \H_0(\gamma), \H(\vec{w})\}$ describe the same dynamics so we have
\begin{equation}\label{reFHs3}
  \F(\H_0)=\F(\H_\phi)=\F(\H_0(\gamma))=\F(\H(\vec{w})) \,,
\end{equation}
for all scalar function $\phi(\vec{\hat{x}})$, positive number $\gamma$ and vector $\vec{w}$.
Note that the Hamiltonians~\eqref{HamilA2}, \eqref{HamilA3b} and \eqref{translq1} can be expressed as
\begin{equation}\label{U1Uphi}
\begin{split}
  &\H_\phi=\U_\phi^\dagger\H_0\U_\phi\,, \\
  &\H_0(\gamma)=\hat{W}(\gamma)^\dagger\H_0\hat{W}(\gamma)\,,\\
  &\H(\vec{w})=\hat{X}^{\dagger}(\vec{w})\H_0\hat{X}(\vec{w})\,,
  \end{split}
\end{equation}
with
\begin{equation*}
\begin{split}
  &\U_\phi=\exp[iq\phi(\vec{\x})]\,, \\
  &\hat{W}(\gamma)=\exp\left[\frac{i}2(\ln\gamma)(\vec{\x}\cdot\vec{\p}+\vec{\p}\cdot\vec{\x})\right]\,, \\
  &\hat{X}(\vec{w})=\exp(i\vec{w}\cdot\vec{\p}~)\,.
  \end{split}
\end{equation*}
The symmetry $\F(H)=\F(\U H\U^\dagger)$ is the simplest and natural way to satisfy physical requirements~\eqref{reFHs3}.




\section{Complexity in a compact Riemannian manifold}
Let us consider a non-relativistic particle with mass $m$ in an $n$-dimensional compact Riemannian manifold $M$ ($\partial M=0$) with arbitrary positive definite metric $G_{\mu\nu}$.
The Hamiltonian $H$ is given by
\begin{equation}\label{defHRiem}
  \H=-\frac{\nabla^2}{2m}+V(x) \,, 
\end{equation}
%
where $\nabla$ is the covariant derivative associated with a metric $G_{\mu\nu}$ and $V(x)$ is a potential bounded below.

As the eigenvalues of such a Hamiltonian cannot be computed analytically in general, we cannot use the $\zeta$-function method. In order to regularize the trace of this Hamiltonian, we define the regularized trace of any function of $\H$, say $f(\H)$, as
\begin{equation}\label{defheatKT}
  \Tr_{\tau}\left[f(\H)\right]:=\frac1{K_\H(\tau)}\Tr\left[f(\H)e^{-\tau \H}\right]\,,
\end{equation}
where
\begin{equation}
K_\H(\tau):=\Tr(e^{-\tau \H})~~~\text{and}~\tau\rightarrow0^+\,.
\end{equation}
The operator $e^{-\tau \H}$ is usually called the ``heat kernel''~\cite{Avramidi:2001ns,Vassilevich:2003xt}. Here, we take $\tau>0$ so that Eq.~\eqref{defheatKT} is finite for the Hamiltonian bounded from below.

First, we compute $\bH$ by using Eq.~\eqref{meanHs1} with the regularized trace \eqref{defheatKT}
\begin{equation}\label{defHeatHmean}
  \bH=\Tr_\tau(\H)/\Tr_\tau(\I)=\Tr_\tau(\H)=-\frac{\td}{\td\tau}\ln K_\H(\tau)\,.
\end{equation}
Thus, we generalize Eq.~\eqref{formforFs1b} as
\begin{equation}  \label{defheatKT40}
\begin{split}
  \F^2(\H)&=\lambda^2\{\Tr_\tau(\H^2)-[\Tr_\tau(\H)]^2\}   \\
  &=\lambda^2\frac{\td^2}{\td \tau^2}\ln K_\H(\tau)\,. 
 \end{split}
\end{equation}
%
For finite dimensional cases, Eq.~\eqref{defheatKT40}  agrees with Eq.~\eqref{formforFs1b} after we take limit $\tau\rightarrow0^+$.

Without loss of generality, we may set $m=1/2$. For a general $G_{\mu\nu}$, it is difficult to find even the ground state and the first eigenvalue of $H$. However, {in the case of $\tau\rightarrow0^+$}, the trace of the heat kernel can be computed in terms of a serie of $\tau$~\cite{Vassilevich:2003xt},
\begin{equation}\label{traceKRiem}
  K_\H(\tau)=\frac1{(4\pi\tau)^{n/2}}\sum_{n=0}^\infty a_n\tau^n\,,
\end{equation}
where $a_0=\int_M\sqrt{G}\td^n x$, $a_1=\int_M(R/6+V)\sqrt{G}\td^n x$ and $a_2=b+c$ with
%
\begin{eqnarray}
  b&:=&\frac1{2}\int_M\sqrt{G}(V+R/6)^2\td^n x\,,  \label{coeffa21}  \\
  c&:=&\frac1{180}\int_M\sqrt{G}(R_{\mu\nu\sigma\rho}R^{\mu\nu\sigma\rho}-R_{\mu\nu}R^{\mu\nu})\td^n x\,. \label{coeffa22}
\end{eqnarray}
Here $G$ is the determinant of $G_{\mu\nu}$, $R_{\mu\nu\sigma\rho}$ is the Riemann tensor, $R_{\mu\nu}$ is the Ricci tensor and $R$ is the scalar curvature.
Plugging Eq.~\eqref{traceKRiem} into Eq.~\eqref{defheatKT40} and setting $\tau\rightarrow0^+$, we obtain
\begin{equation}\label{bareFRiem}
  \lambda^{-2}\F^2(\H)=\frac{n}{2\tau^2}+\frac{2a_2a_0-a_1^2}{a_0^2}+\mathcal{O}(\tau)\,.
\end{equation}
The divergent term  does not contribute to the metric defined by Eq.~\eqref{defmetricg}, so it can be removed without changing the complexity geometry. Thus we obtain the renormalized norm squared
%
\begin{equation}\label{reFRiem}
  \F^2_{\text{re}}(\H):=\lim_{\tau\rightarrow0^+}\F^2(\H)-\frac{\lambda^2n}{2\tau^2}=\lambda^2\frac{2a_2a_0-a_1^2}{a_0^2}\,.
\end{equation}
%
For a ``free particle'', i.e., a flat metric $G_{\mu\nu}=\delta_{\mu\nu}$ with a constant potential $V(x)=V_0$, we obtain $\F^2_{\text{re}}(\H)=0$, which is consistent with \eqref{normfreeH21}.

For the complexity to be well-defined $\F^2_{\text{re}}(\H)$ must be nonnegative but the right hand side of \eqref{reFRiem} may not be nonnegative in general. Let us find the condition for this to be nonnegative. First, the numerator of Eq.~\eqref{reFRiem}, can be rewritten as
\begin{equation} \label{xxxy}
2a_2a_0-a_1^2= \left(2ba_0-a_1^2\right)+2ca_0\,.
\end{equation}
Thanks to the Cauchy-Schwarz inequality
\begin{equation}\label{CSineq}
\begin{split}
  2ba_0&=\left(\int_M(V+R/6)^2\sqrt{G}\td^n x\right)\int_M1^2\sqrt{G}\td^n x\\
  &\geq\left(\int_M(V+R/6)\times1\sqrt{G}\td^n x\right)^2=a_1^2\,,
  \end{split}
\end{equation}
the term in the parenthesis in Eq.~\eqref{xxxy} is nonnegative so we focus on the last term, $2ca_0$.

For a flat space, $c=0$ so Eq.~\eqref{xxxy} is always nonnegative in arbitrary dimension. It can be zero only for free particles, i.e., $V(x)$ is constant. It is consistent with Eq.~\eqref{normfreeH21}.
For a curved space, as $a_0$ is positive, we only need to consider the sign of $c$.
For $n=1$, $R_{\mu\nu\sigma\rho}=0$ so $c=0$. For  $n=2$, $R_{\mu\nu\sigma\rho}$ has only one independent term. Under a local orthonormalized frame $\{e_i\}$ the metric components read $G_{ij}=\delta_{ij}$ and the nonzero curvature component is $R_{1212} =: K$. Thus, $R_{\mu\nu\sigma\rho}R^{\mu\nu\sigma\rho}-R_{\mu\nu}R^{\mu\nu}=2K^2\geq0$ and $c\geq0$. For $n\geq3$, we rewrite $c$ as
\begin{equation}\label{ctermin3d}
   c=c_{\text{GB}}+\frac1{180}\int_M\sqrt{G}(3R_{\mu\nu}R^{\mu\nu}-R^2)\td^3 x\,,
\end{equation}
where the Gauss-Bonnet term is introduced:
\begin{equation}
c_{\text{GB}}:=\frac1{180}\int_M\sqrt{G}(R_{\mu\nu\sigma\rho}R^{\mu\nu\sigma\rho}-4R_{\mu\nu}R^{\mu\nu}+R^2)\td^n x \,.
\end{equation}
For $n=3$, $c_{\text{GB}}=0$. In the local orthonormalized frame $\{e_i\}$, we can diagonalize the Ricci tensor and obtain three eigenvalues $\{k_1,k_2,k_3\}$. Thus, $3R_{\mu\nu}R^{\mu\nu}-R^2=3(k_1^2+k_2^2+k_3^2)-(k_1+k_2+k_3)^2\geq0$ and $c\geq0$.
However, for $n\geq4$, $3R_{\mu\nu}R^{\mu\nu}-R^2$ and $c_{\text{GB}}$ can be negative for some geometries, so $c$ and  Eq.~\eqref{xxxy} can be negative.

The Hamiltonian~\eqref{defHRiem}
can be used in low energy limit of quantum field theory. If the spacetime dimension in low energy limit is 3+1 or less, Eq.~\eqref{reFRiem} is always nonnegative.  This is compatible with the spacetime of our world in low energy limit. It can be also understood as follows. If we assume that a physical Hamiltonian should give a nonnegative complexity, it gives some restriction on the dimension and geometry of spacetime.

%


\section{Conclusions}
For the complexity to be a useful tool for gravity and QFT/QM, we first have to
fill up the conceptual gap between the quantum circuits and QFT/QM in defining complexity. By noting that the generating functional plays a central role in QFT/QM contrary to quantum circuits we propose
an additional symmetry~\eqref{untrary-inv} for the complexity in QFT/QM. It gives a simple and unique formula for the complexity of SU($n$) operators,{ Eqs. \eqref{GbiFs1} and \eqref{CforHs1}.

Even though the formula is unique, its result is still rich when it is generalized for physical Hamiltonians.}  In particular, for the complexity of the operators generated by Hamiltonians in an infinite dimensional Hilbert space,
our complexity formula gives novel results which can not be obtained from finite qubit systems.
Interestingly enough, the complexity geometry corresponding to a general quadratic Hamiltonian in one-dimension is equivalent to AdS$_3$ space.
Here, we pointed out the lower boundedness of the Hamiltonian gives a  constraint for the non-negativity of the complexity, which has not been appreciated before. We want to stress that three basic physics, non-negative complexity, bounded-below Hamiltonian and sub-luminal speed of particles are three aspects of the same thing in our proposal.

Our formula proves that the complexity of the operator generated by free Hamiltonians vanishes, which was intuitively plausible. 
We uncovered the connection between the complexity and the background geometry. In particular, the fact that the critical dimension to ensure a nonnegative complexity in low energy limit is just 3+1 dimension is worthy of more investigations.

We want to stress that the non-negativity of complexity in {\it infinite} dimensional Hilbert spaces  is not trivial, while it is trivial in finite dimensional cases for all Hermitian Hamiltonians. When we generalize our formula~\eqref{formforFs1b} to infinite dimensional Hilbert spaces, the results are still nonnegative but divergent, so regularization and renormalization are needed. Then we find that the renormalized complexity is no longer always non-negative if we do not make any restriction on the Hamiltonian. For two models considered in this paper, we found that the requirement of non-negativity can lead to some requirements so that the underlying models are physically correct. 
For quadratic Hamiltonians, it implies that the Hamiltonian should be bounded below; for a non-relativistic particle, it implies  that the dimension in low energy limit should be less than 3+1. These results seem to propose a new universal principle: the Hamiltonian of a physically correct theory should give us nonnegative complexity.

This paper focused on a few quantum mechanical systems in order to simplify technical details. We think that our proposal can be applied to QFT in principle and may result in more novel properties in QFT, especially at strong interactions or in the curved spacetime. However, we would like to mention that this generalization may not be simple technically. For example, for a free scalar field theory, there are two kinds of infinity: the infinite degrees of freedom and the infinite Hilbert space of every single degree of freedom. The methods in this paper can deal with the latter but cannot cover the former, especially, when we consider some non-perturbative field theories. Thus, some new renormalization techniques may be necessary, which we would like to address in future.

\acknowledgments
We would like to thank Yu-Sen An for helpful discussions during this work. The work of K.-Y. Kim was supported by Basic Science Research Program through the National Research Foundation of Korea(NRF) funded by the Ministry of Science, ICT $\&$ Future Planning(NRF- 2017R1A2B4004810) and GIST Research Institute(GRI) grant funded by the GIST in 2018. We also would like to thank the APCTP(Asia-Pacific Center for Theoretical Physics) focus program,``Holography and geometry of quantum entanglement'' in Seoul, Korea for the hospitality during our visit, where part of this work was done.

\appendix

\section{{Importance of bi-invariance for the complexity in QFT/QM}} \label{unitrary1}

In this appendix, we explain why the bi-invariance is important for the complexity in QFT/QM. There are various arguments to support Eq~\eqref{untrary-inv} and bi-invariacne in Refs.~\cite{Yang:2018nda,Yang:20180919} and section \ref{uni123}. Among many discussion in Refs.~\cite{Yang:2018nda,Yang:20180919}, here we introduce only two arguments briefly.

Before starting, we would like to emphasize again that the only right-invariance is perfectly fine for quantum circuits. What we want to highlight is that there is something more to consider for general QFT/QM systems which are different from quantum circuits.

\subsection{Right-invariance or left-invariance?}
In Nielsen's and many other related works, the complexity is claimed to be right-invariant based on the following argument~\cite{Brown:2017jil}.
Let us consider the construction of $\U$  from $\W$ by a series of gates $\{g_1,g_2,\cdots,g_N\}$ in a time order (right to left):
\begin{equation}\label{rightgates1}
  \U=g_Ng_{N-1}\cdots g_2g_1\W\,.
\end{equation}
One may define the ``relative complexity'' $d_r(\U,\W)$ by the minimal number of gates to construct the operator $\U$ from the operator $\W$. Here, we added the index ``$r$'' to express that the old operator will appear at the right-side of the product (i.e. `r'ight to left).  For the initial operator $\W=\I$ the relative complexity becomes just the complexity.
For a right-translation: $\U\rightarrow\U\x$ and $\W\rightarrow\W\x$, 
\begin{equation}\label{rightgates2}
  \U\x=g_Ng_{N-1}\cdots g_2g_1\W\x\,,
\end{equation}
%
it is obvious
%
\begin{equation}
d_r(\U,\W)=d_r(\U\x,\W\x)\,,
\end{equation}
%
which means that the (relative) complexity should be right-invariant.

If the product order in Eq.~\eqref{rightgates1} is the only way for the time order this argument works. However, there is \textit{no physical, mathematical} or \textit{logical} reason to forbid us from choosing the `left to right' product order as a time order
\begin{equation}\label{rightgatesb1}
  \U=\W g_1'g_{2}'\cdots g_{K-1}'g_K'\,.
\end{equation}
In this case, it is also obvious that
\begin{equation}
d_l(\U,\W)=d_l(\x\U,\x\W)\,,
\end{equation}
where the index ``$l$'' means that the old operator will appear at the left-side of the product (i.e. `l'eft to right). Thus, the complexity should be left-invariant.

Indeed, this kind of ordering choice is familiar also in quantum mechanics.
The Schr\"{o}dinger's equation is written as
\begin{equation}\label{Scheq1}
  \frac{\td}{\td t}|\psi(t)\rangle=iH(t)|\psi(t)\rangle\,,
\end{equation}
which implies that the time evolution operator is
\begin{equation}\label{timeUt1}
  \overleftarrow{\cal{P}}\exp\left[\int_0^t i H(s) \td s\right]\,.
\end{equation}
With this ket $|\cdot\rangle$ `representation' the product order~\eqref{rightgates1} stands for the time order. However, equivalently, we may use  a bra $\langle\cdot|$ to represent a quantum state without changing any physics. In this case, the Schr\"{o}dinger's equation reads
\begin{equation}\label{Scheq2}
  \frac{\td}{\td t}\langle\psi(t)|=-i\langle\psi(t)|H(t)\,,
\end{equation}
which implies that the time evolution operator is
\begin{equation}\label{timeUt2}
  \overrightarrow{\cal{P}}\exp\left[-\int_0^t i H(s) \td s\right]\,,
\end{equation}
and the product order~\eqref{rightgatesb1} stands for the time order.


However, there is an important difference between the above two examples: the complexity and quantum mechanics.
For quantum mechanics, it does not matter if we choose `left to right' product or `right to left' product; this choice is completely a convention and all physics are the same regardless of this choice. For complexity, it matters because in general
\begin{equation} \label{diff1}
d_r(\U,\W)\neq d_l(\U,\W)\,.
\end{equation}
Is this `ordering-dependence' for complexity problematic? The answer is no and yes.

For the complexity {\it in quantum circuit}  the `ordering-dependence' is perfectly fine and even natural.  In this case, the rules to construct circuits are made by human and one can choose either Eq.~\eqref{rightgates1} or Eq.~\eqref{rightgatesb1}. The complexity is defined based on this choice, so there is nothing wrong with Eq. \eqref{diff1}. Nielsen simply chose Eq.~\eqref{rightgates1} for his {\it quantum circuit} construction so his complexity must be right-invariant.
If Nielsen have chosen Eq.~\eqref{rightgatesb1} all of his results are still valid now with a {\it left-invariant} complexity.

However, when we generalize the concept of complexity in quantum circuit to {\it QFT/QM}, we (human) are not the one to construct unitary operations (the counterpart of the quantum circuit);  there is no reason to favor between Eq. \eqref{timeUt1} and Eq. \eqref{timeUt2}. Furthermore, Eq. \eqref{diff1} is not satisfactory because it means a physical quantity depends on convention.
Therefore, it will be natural to impose both right and left invariance at the same time (bi-invariance). Furthermore, it turns out that this bi-invariance implies~\cite{Yang:2018nda,Yang:20180919}
\begin{equation} \label{diff11}
d_r(\U,\W) = d_l(\U,\W)\,,
\end{equation}
which is more natural from the {\it QFT/QM} perspective.

\subsection{Compatibility with the QFT/QM framework}
There is another issue with the {\it only} right or left complexity; it may not be compatible with the framework of QFT/QM.

Let us first recall that a physical Hamiltonian $H$ and $U H U^\dagger$ (with a unitary operator $U$) are equivalent in QFT in the sense that their  generating functionals and relevant physics are the same. The question now is if the complexity generated by $H$ and $U H U^\dagger$ are the same or not. As we have shown at the end of Sec.~\ref{overviweC}, if the complexity is only right (or left)-invariant, it implies $\F(H)\neq \F(\U H\U^\dagger)$ in general. Thus, the complexities of two evolutions generated by $H$ and $\U H\U^\dagger$ may be different. Does this mean that the complexity is an observable which can distinguish the system of $H$ from $U H U^\dagger$? In principle, the answer can be `yes'; it is possible that the generating functional may not have complete information on the complexity ({for example, quantum circuits}). However, in our current framework of QFT, {it is assumed the generating functional contains all the observable information of the QFT so }
it will be more natural to impose $\F(H) = \F(\U H\U^\dagger)$ if the complexity is an observable. This amounts to the bi-invariance of the complexity.

The compatibility of $\F(H) = \F(\U H\U^\dagger)$ with QFT/QM can be shown more explicitly if we recall that there are \textit{infinitely many different} Hamiltonians describing the same system in QFT/QM.
For example, let us consider a following Lagrangian of a particle.
\begin{equation}\label{defLxxt1}
  L=L(x,\dot{x},t)\,.
\end{equation}
The Hamiltonian is given by a Lagendre transformation
\begin{equation}\label{defL2H1}
  \H=\H(x,p,t):=\dot{x}\frac{\partial L}{\partial\dot{x}}-L(x,\dot{x},t),
\end{equation}
with $\dot{x}$ is a function of $x,p$ and $t$. However, for a given physical system, the Lagrangian is not unique. We can define an new but {\it physically equivalent} Lagrangian $\tilde{L}$ such that
\begin{equation}\label{defLxxt2}
  \tilde{L}=L(x,\dot{x},t)+\phi'(x)\dot{x}\,,
\end{equation}
where $\phi'(x):=\frac{\td\phi(x)}{\td x}$ and $\phi(x)$ is arbitrary smooth function of $x$.
It can give us a new Hamiltonian
\begin{equation}\label{defL2H1}
  \tilde{\H}=\dot{x}\frac{\partial\tilde{L}}{\partial\dot{x}}-\tilde{L}(x,\dot{x},t)=\H(x,p-\phi',t)=e^{i\phi(x)}\H e^{-i\phi(x)}\,,
\end{equation}
where $\dot{x}$ is a function of $x,p$ and $t$ by $p=\partial\tilde{L}/\partial\dot{x}=\partial L/\partial\dot{x}+\phi'$.\footnote{For example, let us consider a harmonic oscillator $L=\frac12m\dot{x}^2-\frac12kx^2$ and $\H(x,p,t)=p^2/(2m)+kx^2/2$. This harmonic oscillator can be described also by a Lagrangian $\tilde{L}=\frac12m\dot{x}^2-\frac12kx^2+\phi'\dot{x}$, and the Hamiltonian reads $\tilde{\H}(x,p,t)=(p-\phi')^2/(2m)+kx^2/2=\H(x,p-\phi',t)$. }

We now may ask the following interesting question. Are the complexities of the operators generated by $\H$ and $\tilde{\H}$ the same or not? In other words, is $\F(\H)$ the same as $\F(e^{i\phi(x)}\H e^{-i\phi(x)})$ or not? If the complexity is bi-invariant $\F(\H) = \F(e^{i\phi(x)}\H e^{-i\phi(x)})$, while
if the complexity is {\it only} right (or left)-invariant,  $\F(\H)\neq\F(e^{i\phi(x)}\H e^{-i\phi(x)})$ in general. However, $\F(\H)\neq\F(e^{i\phi(x)}\H e^{-i\phi(x)})$ implies that two Lagrangians~\eqref{defLxxt1} and \eqref{defLxxt2} are {\it not physically equivalent}, which challenges the current frameworks of QFT/QM. Thus, it is very natural to impose the condition $\F(H)=\F(\U H\U^\dagger)$.
This issue has been addressed in Ref.~\cite{Yang:20180919} in more detail and it has been shown that for arbitrary Hermit operator $H(x,p)$, which is smooth and have a Taylor's expansion with respective to $p$ at $p=0$, the equivalence of $L$ and $\tilde{L}$ implies
\begin{equation}\label{unitraryH1}
  \F(\H)=\F(e^{iH(x,p)}\H e^{-iH(x,p)})\,.
\end{equation}

Here, we would like to emphasize again that such a compatibility issue appears only when we develop the Nielsen's complexity theory in {\it QFT/QM}  not in {\it quantum circuits}.  Quantum circuits are not Lagrangian or Hamiltonian dynamical systems so the arguments between Eq.~\eqref{defLxxt1} and Eq.~\eqref{unitraryH1} do not make sense for quantum circuits. In quantum circuits, the generating functional does not play a crucial role and the symmetry $\F(H)=\F(\U H\U^\dagger)$ may not be necessary from this perspective. In fact, in a quantum circuit, the ``gate'' is a real physical object realizing abstract unitary operation. The unitary transformation of the gate yields a different gate or a product of gates. Thus, it is possible that the cost (number of gates) of a given operator is not invariant under a unitary transformation, i.e. $\F(H)\ne\F(\U H\U^\dagger)$.

\subsection{More technical details for the unitary invariance}

What if the complexity is not an observable but a physical quantity yielding observables? For example, a wavefunction of the Shr\"{o}dinger's equation and the gauge potential in electromagnetism are such quantities. Here, even in this case, we will show that $\F(H)=\F(\U H\U^\dagger)$ is still required.

Let us first consider the case that the complexity itself leads to an observable. This means that the norm $\F$ can lead to an observable, i.e., an observable $O(\F)$ as a function of $\F$. Indeed, this one variable case is trivial. In order for $O$ to be an observable $\F$ must be an observable since $\F$ is a scalar.   Notwithstanding, we describe it for completeness.
We assume that $O(x)$ is a non-constant and smooth function, which means
\begin{equation}\label{reqnonconst1}
  \forall x,~~~\exists k\in \mathbf{N}^+\,,~~~\text{such that}~\frac{\td^k O(x)}{\td x^k} \neq0 \,.
\end{equation}
Because $O$ is an observable, for arbitrary $H$ and $\U(s)$,
\begin{equation}\label{proofeq11}
  O(\F(H)) =O(\F(\U(s) H\U^\dagger(s)))\  \Leftrightarrow \ \frac{\td O}{\td s} =0 \,,
\end{equation}
where $s$ is a continuous parameter for $\hat{U}(s)$.  If $\td O(x) / \td x \ne 0 $ Eq. \eqref{proofeq11} implies
\begin{equation}\label{eqforHUs1}
  \frac{\td}{\td s}\F(\U(s) H\U(s)^\dagger)=0\,,
\end{equation}
because
\begin{equation}\label{dfsds1}
 \frac{\td O}{\td s } =\frac{\td O(x)}{\td x} \frac{\td}{\td s}\F(\U(s)  H\U(s)^\dagger) =0  \,.
\end{equation}
If $\td O(x) / \td x = 0 $, we can find a $k \ge 2$ such that $\td^k O(x) / \td x^k \neq 0$ (Eq. \eqref{reqnonconst1}) and obtain the same result Eq. \eqref{eqforHUs1}. This shows $\F(H)=\F(\U H\U^\dagger)$ for arbitrary $\U$.

In more general cases, one complexity itself may not be an observable but we can obtain an observable by several complexities. For example, one may think that the complexity is not  a direct observable but the difference between two complexities is an observable, which implies that neither $\F(H_1)$ nor $\F(H_2)$ is an observable but $\F(H_1)-\F(H_2)$ is an observable. Thus we have $\F(H_1)-\F(H_2)=\F(\U H_1\U^{\dagger})-\F(\U H_2\U^{\dagger})$. Then we can choose $O:=O(x_1,x_2)=x_1-x_2$ and so $O(\F(H_1),\F(H_2))=O(\F(\U H_1\U^\dagger),\F(\U H_2\U^\dagger))$.
In general an observable $O$ may have more than two arguments.

If  $O(x_1,x_2,\cdots,x_n)$ is a non-constant smooth function, we can find that $\forall x_1,x_2,\cdots,x_n$, $\exists k\in \mathbf{N}^+$ and $\exists l$ such that $\partial^k O/\partial x_l^k\neq0$. Without loss of generality, we take $l=1$ and obtain
\begin{equation}\label{reqnonconst}
  \exists k\in \mathbf{N}^+,~~~\partial^k O/\partial x_1^k\neq0,~~~\forall x_1,x_2,\cdots,x_n\,.
\end{equation}
Next, as $O(\F(H_1),\F(H_2),\cdots,\F(H_n))$ is an observable, we require:
\begin{equation}\label{proofeq10}
  \begin{split}
  &O(\F(H_1),\F(H_2),\cdots,\F(H_n))\\
  =&O(\F(\U H_1\U^\dagger),\F(\U H_2\U^\dagger),\cdots,\F(\U H_n\U^\dagger)) \,,
  \end{split}
\end{equation}
for arbitrary $H_j$ and $\U$. i.e. $H_j$ and $\U H_j\U^\dagger$ gives the same observables. The basic idea of the proof is to reduce Eq.~\eqref{proofeq10} to the trivial one variable case.
\begin{equation}\label{proofeq2}
  \begin{split}
  &O(\F(H),\F(H'),\cdots,\F(H'))\\
=&O(\F(\U(s) H\U(s)^\dagger),\F(H'),\cdots,\F(H'))\,.
  \end{split}
\end{equation}
It can be done by choosing $H_1=H, H_2=H_3=\cdots H_n=H'$ and $U(s)=\exp(iH's)$  with $\U(s) H\U(s)^\dagger\neq H$. By the essentially same procedure as the one variable case, we conclude $\F(H)=\F(\U H\U^\dagger)$.

\section{Complexity metric for SU(1,1) group}\label{app2}
We will explain how to obtain  Eq.~\eqref{zetametricSU11}.
Let us start with Eq.~\eqref{defRH1},
%
\begin{equation}\label{decomp11}
  \td\U(X^I)\U^\dagger(X^I)=ie_IY^I(s)\td s\,,
\end{equation}
where $U(X^I)$ is an element of the group $G$ parameterized by  the coordinate $X^I$ and the Lie algebra $\mathfrak{g}$ is spanned by the bases $\{ie_I\}$.

For any representation of $G$, i.e. a map $\pi: G\mapsto$GL($n,K$) with $K=\mathbb{R}$ or $\mathbb{C}$, and its induced representation in  the Lie algebra, i.e. $\pi_*: \mathfrak{g}\mapsto$GL($n,K$) Eq.~\eqref{decomp11} reads
\begin{equation}\label{decomp10}
\begin{split}
  \td[\pi(\U)][\pi(\U)]^{-1}&=\pi_*(ie_I)Y^I(s)\td s\,.
  \end{split}
\end{equation}
Because the coefficient $Y^I\td s$ is independent of the choice of faithful representations we may choose any faithful representation for our convenience.

For example, we may parameterize SU(1,1) group as
\begin{equation}\label{parameterU112}
  \U(X^I) = \U(y, z,u)=\exp(iy e_0)\exp(iz e_+)\exp(iue_-)\,,
\end{equation}
and choose a $2\times 2$ matrix representation for $e_0$ and $e_\pm$ as follows. $\pi_*(ie_I)=K_I$ with
\begin{equation*}
  K_0=\frac12\begin{pmatrix}
  1&0\\
  0&-1
  \end{pmatrix},~~
  K_+=\begin{pmatrix}
  0&-1\\
  0&0
  \end{pmatrix},~~K_-:=\begin{pmatrix}
  0&0\\
  1&0
  \end{pmatrix}\,.
\end{equation*}
In this representation the group element Eq.~\eqref{parameterU112} is expressed as
\begin{equation}\label{repUyzu1}
  \pi(\U(y,z,u))=\begin{pmatrix}
  e^{y/2}(1-zu),&-e^{y/2}z\\
  e^{-y/2}u,&e^{-y/2}
  \end{pmatrix} \,,
\end{equation}
which yields
\begin{equation}\label{decomp2}
\begin{split}
  \td[\pi(\U)]\pi(\U)^{-1}=&(\td y-2z\td u)K_0+e^y(\td z+z^2\td u)K_+\\
  &+e^{-y}\td uK_-\,.
  \end{split}
\end{equation}
%
Thus, by Eq. \eqref{decomp10} we find
\begin{equation}\label{valueYrYlB1a}
  Y^I\td s=\left(\td y-2z\td u, e^y(\td z+z^2\td u), e^{-y}\td u\right) \,.
\end{equation}
%

Finally, the line element reads
%
\begin{equation}\label{metricYpm0}
 \td l^2=\tilde{g}_{IJ}Y^I Y^J\td s^2
 =\lambda_0^2(Y^+Y^--(Y^0)^2/4)\td s^2\,.
\end{equation}
where $\tilde{g}_{IJ}$ is given by Eq.~\eqref{F211}
and \eqref{defmetricg}.
By Eq.~\eqref{valueYrYlB1a} we obtain the metric in the group manifold
\begin{equation}\label{zetametricSU111}
  \td l^2=\lambda_0^2(-\td y^2/4+z\td y\td u+\td z\td u)\,.
\end{equation}

\bibliographystyle{JHEP}

\providecommand{\href}[2]{#2}\begingroup\raggedright\endgroup

\end{document}